\documentclass[11pt,a4paper]{article}
\usepackage[hyperref]{tto2022}
\usepackage{times}
\usepackage{latexsym}
\usepackage{graphicx}
\usepackage{tipa}
\usepackage[T1]{fontenc}
\usepackage{hyperref}
\usepackage{url}

\aclfinalcopy % Uncomment this line for the final submission

\title{YouTube COVID-19 Vaccine Misinformation on Twitter: Platform Interactions and Moderation Blind Spots}

\author{  David S. Axelrod \\
  Indiana University \\
  Bloomington \\
  Luddy SICE \\
  \texttt{daaxelro@iu.edu} \\\And
  Brian P. Harper \\
  Indiana University \\
  Bloomington \\
  Luddy SICE \\
  \texttt{bpharper@iu.edu} \\\And 
  John C. Paolillo \\
  Indiana University \\
  Bloomington \\
  Luddy SICE \\
  \texttt{paolillo@indiana.edu} \\
}

\date{}

\begin{document}
\maketitle
\begin{abstract}
  While most social media companies have attempted to address the challenge of COVID-19 misinformation, the success of those policies is difficult to assess, especially when focusing on individual platforms. This study explores the relationship between Twitter and YouTube in spreading COVID-19 vaccine-related misinformation through a mixed-methods approach to analyzing a collection of tweets in 2021 sharing YouTube videos where those Twitter accounts had also linked to deleted YouTube videos. Principal components, cluster and network analyses are used to group the videos and tweets into interpretable groups by shared tweet dates, terms and sharing patterns; content analysis is employed to assess the orientation of tweets and videos to COVID-19 messages. From this we observe that a preponderance of anti-vaccine messaging remains among users who previously shared suspect information, in which a dissident political framing dominates, and which suggests moderation policy inefficacy where the platforms interact.
\end{abstract}

\section{Introduction}
\label{sec:intro}
During the COVID-19 pandemic, Social media platforms acted rapidly to staunch misinformation on their platforms. This has led to policies addressing misinformation that are inconsistent internally and between each other\cite{krishnan2021research}. Twitter updated its COVID-19 misinformation policy at the beginning of 2021 and throughout the year to address vaccine misinformation \cite{twitter_mispol}. YouTube's COVID-19 policies were similar, but also added a separate vaccine misinformation policy in late September\cite{youtube_vaccinepress,youtube_mispol}. Though different in format and exceptions, both policies obligate the respective platforms to moderate vaccine misinformation.

While misinformation is regularly deleted on both platforms, new misinforming content appears in a back-and-forth process from which external researchers may only catch glimpses. Assessing the quality and efficacy of anti-misinformation policy is therefore quite difficult, not least because of interactions between different social media platforms and how they may amplify each others' misinformation. While platform policies and their implementation are centered around the holdings of a specific corporation, users may use multiple platforms irrespective or ownership, and so may readily exploit inter-platform differences.

What then is the shared role of Twitter and YouTube in the spread of vaccine misinformation? How do the interfaces between the two platforms and their policy differences contribute to it? And what responsibility do users bear in the process of circulating such misinformation? This paper addresses these questions through a mixed-methods analysis of a set of tweets sharing video links to likely vaccine misinformation.

\section{Background}
\label{sec:back}
Previous studies have suggested a link between social media misinformation and vaccine hesitancy behaviors \cite{loomba2021measuring,pierri2022online}, and while there may be reason to be cautious about proposing causal links \cite{valensise2021lack}, it is clear that an anti-vaccine echo chamber resides among social media. COVID-19 research on YouTube has often attempted to assess whether or not its content is dominated by misinformation, with results generally indicating that news coverage predominated over misinformation on YouTube in 2020 \cite{knuutila2020covid,paolillo2022covid,marchal2020covid19,andika2021youtube,li2020youtube}, though misinformation was always present. Other studies assess the quality of medical communication on the YouTube platform, finding mixtures of good and bad information quality \cite{szmuda2020youtube,basch2020preventive}. More recently, research has shifted from addressing the pandemic itself to assessing the state of vaccine messaging and its role in combating misinformation \cite{jennings2021lack,laforet2022understanding}. The situation is broadly similar for research on Twitter, with early attempts to identify misinformation about the virus and the pandemic \cite{gallotti2020assessing,mourad2020critical} shifting toward vaccine misinformation \cite{yousefinaghani2021analysis}, often assessing the role of bots and the presence of active anti-vaccine campaigns \cite{shi2020social,sharma2022covid}.

Rather than treat different platforms as separate entities, it is better to conceptualize them as parts of a platform ecosystem \cite{van2018platform}, and cross-platform work is important for understanding both how platform systems interact with each other and how users experience individual platforms. Unfortunately, while good cross-platform work has been done \cite{cinelli2020covid}, platform dyads contain individual patterns of interaction that are not necessarily visible on a grand scale. With respect to YouTube and Twitter, links to YouTube videos within COVID-19 tweets have been previously observed \cite{yang2021covid}, and Ginossar et al.~\shortcite{ginossar2022cross} found evidence to suggest that Twitter links to YouTube videos were effective means of spreading misinformation during the first half of 2020, often using prior conspiratorial or anti-vaccine content on YouTube to do so. The present paper continues the work of observing this cross-platform interaction, albeit with different methods of data collection and analysis. 

\begin{figure*}[!t]
	\centering
		\includegraphics[scale=.12]{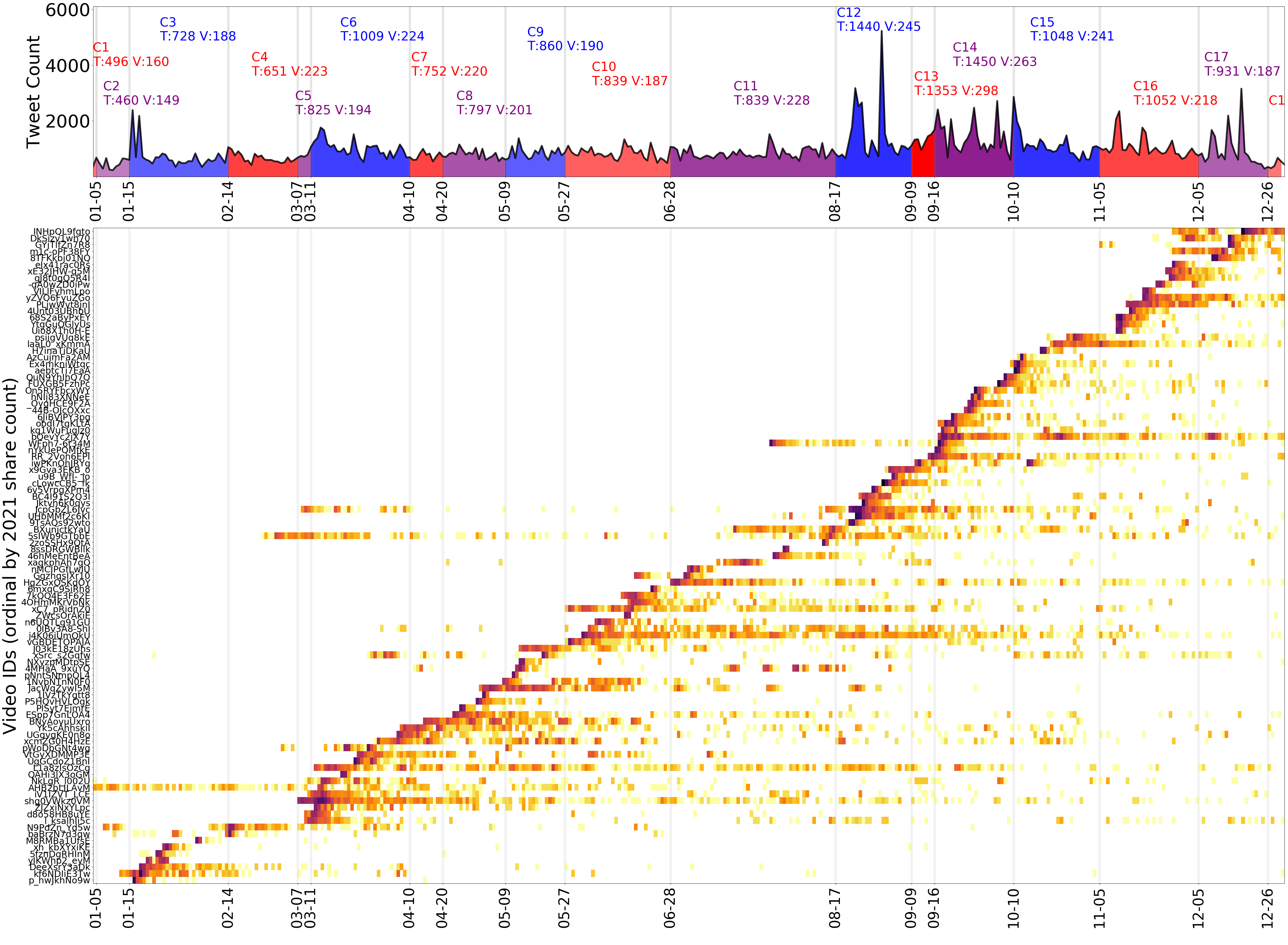}
	\caption{Top: Time series for tweets linking to YouTube videos in our dataset. The time series is segmented into date groupings (see section \ref{sec:dates}), labeled with cluster identities and counts of the tweets and videos in each. Dates shown are the last day in each date cluster. Vertical lines extend down into the heat map to guide the eye. Bottom: Heat map of 100 most shared videos (2021). Heat map color encodes daily frequency for each video. Left-hand y-axis lists video IDs ordered by the date of their maximum daily share count.  }
	\label{fig:hm_ts_grid_3}
\end{figure*}

\section{Methods}
\label{sec:meth}
To address our questions, we conducted an exploratory analysis around a dataset likely to contain vaccine misinformation collected at the interface of Twitter and YouTube. This interface is hosted almost entirely on Twitter: while active YouTubers often have Twitter accounts and link to them, Twitter's feature of shortening and including links allows YouTube videos to be shared in discourses they would not otherwise reach. Twitter's public data API allows one to search tweets for a given time period, possibly filtering them for sites they link to, and YouTube's API allows one to ascertain if a video link still exists on the platform. Through these means, it is possible to construct a view of the YouTube-Twitter platform interface that reflects interaction around COVID-19 vaccine misinformation. 

Following acquisition of a relevant dataset, two general research methods are available: quantitative exploration (e.g., via cluster analysis and network analysis), and qualitative analysis. The approach followed here is to employ different stages of quantitative and qualitative analysis to support each other. This is necessary as the data collected are of a substantial scale that is difficult to approach, and only systematic examination of the videos and tweet texts by researchers can tell us what they actually mean for COVID-19 vaccine discourse.

Since our dataset covers a broad time period (Jan 4 to Dec 31, 2021), the state of information around COVID-19 vaccines and treatment changes greatly during the sample period, meaning it is likely that patterns of video production on YouTube and/or sharing on Twitter also change. Hence, our first step is to organize the time period into approachable groupings of dates based on the video tweeting patterns over time. This information was then presented in the form of a web interface providing links to the relevant videos alongside the tweets from the database; after viewing selected videos, we could then make qualitative judgments in the interface regarding their content. These judgments were then analyzed for consistency and employed in subsequent network analyses. An additional set of analyses were conducted on the term distribution in the tweets; these did not reveal vaccine-related topics as had been hoped, but rather formulaic patterns used for marketing and spam, which were prevalent in our dataset. 

Our tweets come from the CoVaxxy project, an effort that collected IDs for vaccine related tweets using almost 80 keywords ~\cite{deverna2021covaxxy} through Twitter's \texttt{statuses/filter} v1.1 API endpoint.\footnote{\url{https://developer.twitter.com/en/docs/twitter-api/v1/tweets/filter-realtime/overview}} In order to examine how YouTube videos are shared in Twitter content that deals with vaccines we first retrieved the full Twitter data and metadata for each tweet, and then identified the URLs embedded in them, extracted unique video identifiers, and queried the YouTube API for the video status using the \texttt{videos:list} endpoint. Since previous research showed that inaccessible videos contain a high proportion of anti-vaccine content such as the ``Plandemic'' conspiracy documentary~\cite{yang2021covid}, we treat any videos removed by Twitter and publicly unavailable as suspicious. Overall, the fraction of vaccine-related tweets linking to YouTube videos in the CoVaxxy dataset was relatively small, with a daily median of 0.52\%. However, among these links, a daily median of 10.95\% were to inaccessible videos. Over the year, there was a decreasing trend in inaccessible videos with a peak of 45\% in July. Because it is estimated that it takes an average of 41 days for YouTube to remove videos violating their terms ~\cite{knuutila2020covid}, we checked the status of videos at least 2 months after the video was last posted on Twitter. Since unavailable content is not available for full investigation, we focus on related but available content by selecting tweets with available videos from users who also shared one of the unavailable videos in our dataset.  

\section{Grouping dates in the sample}
\label{sec:dates}
The sampling procedure described above resulted in a set of 339,763 tweets (126,244 original and 213,519 retweets) containing 34,819 distinct YouTube video links; this is the cross-platform dataset we seek to explore. Our first question concerns whether there are discernible video tweeting patterns over time. We approached this by examining the timestamps of tweets sharing a common video; this is visualized as a heat map of the 100 most viewed videos in Figure \ref{fig:hm_ts_grid_3}, in which the diagonal pattern clearly indicates that each video is shared on or close to a specific date (generally close to its publication date, though there are exceptions), while in some cases a horizontal dashed line of points indicates that a video might be shared over some longer set of dates, not always close together in time.

\begin{figure*}[!htbp]
	\centering
		\includegraphics[scale=.54]{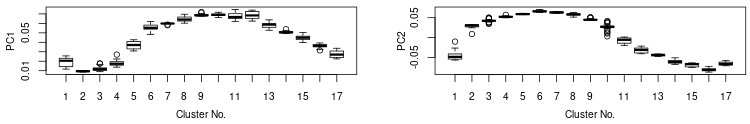}
	\caption{Boxplots of the date groupings on PC1 and PC2, showing each resulting date range to be distinct in terms of video sharing from the others in the sample.}
	\label{fig:boxplotsmall_3}
\end{figure*}

To extract this pattern, a PCA was conducted on the video-date incidence matrix, which yielded three potentially usable dimensions accounting for 20.0\% of the shared variance. This was followed by cluster analysis of the dates on the PCs using the R package mclust \cite{scrucca2016mclust}; this approach permits us to evaluate a broad range of cluster solutions within a principled, model-based framework. A 17-cluster variable volume, shape and orientation (VVV) solution was selected as having optimal BIC; the clusters break the sequential dates into 17 contiguous groups, with the exception of cluster 1, which groups together the earliest and latest dates on account of having fewer videos shared with other dates (Figure \ref{fig:boxplotsmall_3}). Hence, date clusters other than 1 suggest epochs within the Twitter sharing of vaccine-related content from YouTube either as periods of common activity or activity lapses between topics with concentrated attention.

\section{Video Coding}
\label{sec:vcoding}
To probe the nature of the date groupings, a content analysis interface was constructed in which a cluster and a date within the cluster could be chosen to present a list of videos to examine, as well as to provide access to the tweets sharing the video. To limit the coding work to manageable levels while focusing on important videos, we committed to coding only those videos tweeted or retweeted at least 5 times on any given day in 2021. Three coders participated in coding the sample, each coding alternating clusters throughout the year. In total, 5,201 videos out of the total 34,819 videos were coded. Observed thematic continuity and overlap across date clusters ensured that coders had a chance to view and discuss the full range of content in our dataset.

Five binary coding check boxes were displayed on each video, to indicate whether it had vaccine-related content \textbf{v}, whether it expressed positive and/or negative valence messages toward vaccination \textbf{p} and \textbf{n}, whether it had unusual characteristics potentially requiring discussion among the coders \textbf{q}, and whether the video possessed a non-publisher context panel \textbf{cx}. For coding \textbf{v}, we accepted ancillary topics like discussions of vaccine mandates and vaccine procedures whether or not claims about the efficacy of vaccination were directly made. The valance of videos \textbf{p} and \textbf{n} was understood to mean what impression of vaccination the video presents. This could be simultaneously positive and negative, as was common in the context of videos about vaccine debates. The fourth category \textbf{q} is intended for potential future work with this dataset and so may be ignored for the present. The fifth category \textbf{cx} assessed the presence of context panels --- in the YouTube player interface, context bars are provided for user information such as regularly misinformed topics, and we wished to assess how consistently YouTube flagged misinforming videos. YouTube also uses context panels to inform users about state-owned media, but we ignored these cases, as they do not relate to potentially misinforming content so much as the quality of sources. In contrast to the other codes, little interpretation is required for coding \textbf{cx}.

\begin{figure}[htbp]
	\centering
		\includegraphics[scale=.45]{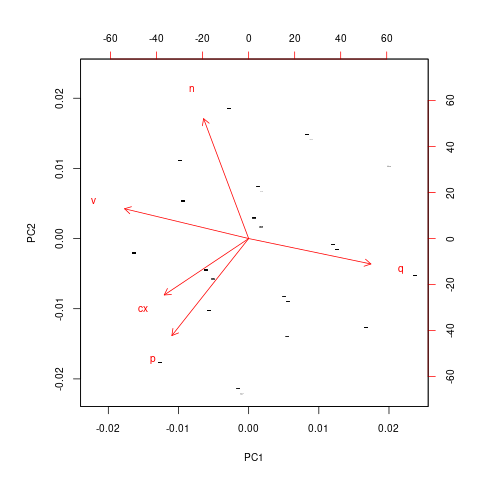}
	\caption{Biplot of the PCA for our coding results. Lines indicate association from videos with our codes. }
	\label{fig:codespca}
\end{figure}

The content coding was analyzed using a combination of PCA and linear modeling. This permitted us to identify patterns of correlation among the codes while simultaneously verifying consistency across coders. We conducted a centered, scaled PCA on the five video codes for the 5201 videos. Two PCs with variance exceeding 1 were retained: PC1 accounts for 51.5\% of the total variation, whereas PC2 accounts for 32.3\% (total 83.9\%, residual 16.1\%). As can be seen in Figure \ref{fig:codespca}, on PC1, in order of decreasing strength, \textbf{v} (vaccine content), \textbf{cx} (context panel), \textbf{p} (positive valence), and \textbf{n} (negative valence) are shifted left (negative), whereas \textbf{q} (flag for further discussion) is on the extreme right (positive) end. This suggests that coding q is negatively correlated with vaccine-related content: as coders found more content there relevant to potential future discussions. PC2 separates the \textbf{n-p} dimension, though \textbf{cx} is loaded reasonably close to \textbf{p}. The lower proportion of shared variation suggests that \textbf{n} and \textbf{p} were less strongly inversely related than \textbf{q} and \textbf{v}. The loading of \textbf{cx} suggests that vaccine-related content was likely to bear a YouTube context panel, but most especially when a positive attitude is expressed than negative; both are otherwise less strongly correlated with \textbf{v} than \textbf{cx}. Hence, the coded videos are characterized by two dimensions: vaccine versus non-vaccine content, and positive or negative valence. Valence is more strongly associated with vaccine-related content, with positive and negative tending to exclude each other and context panels tending to appear on positive valence videos.

\subsection{Inter-rater reliability}
\label{ssec:irr}
Linear model (least squares) regressions of PC1 and PC2 with respect to coder were conducted to test for inter-coder reliability. So long as the contribution of coder to the explained variance of the regression is small, we can be assured that coders are consistent. As the PCs are orthogonal, they are tested individually; while coder is significant in both, the N is large and the proportion of variance accounted for by coder is very small (the adjusted R-squared is 0.9\% for PC1, for PC2 it is 1.96\%). So that we may readily inspect the inter-coder differences, we use sum contrasts (Table \ref{conditmeans}) and report the conditional means for each coder, centered on zero. For PC1, only the positive parameter is significant, meaning there is a significant difference between coders on either side of zero, but not between the two on the negative side. From this, it appears that coder 2 may have greater use of \textbf{q} or less use of \textbf{v} than the other coders. For PC2 both coder parameters are significant; while coder 3 is close to coder 2 in value, they appear to differ in use of the valence codes. Coder 1 appears to have significantly more \textbf{p}/\textbf{cx} than the other two. This could be a difference in either the coding of valence or in the way that YouTube has handled context panels across the different dates of our sample.

\begin{table}[h!]
  \begin{center}
    \label{tab:tablex}
    \begin{tabular}{crrr} 
      \textbf{PC} & \textbf{Coder1} & \textbf{Coder2} & \textbf{Coder3}\\
      \hline
      1 & -0.0446 &  0.2406 & -0.1960 \\
      2 & 0.2371 & -0.1484 & -0.0887 \\
    \end{tabular}
  \end{center}
  \caption{\label{conditmeans}Conditional means for coder models of PCs}
\end{table}

As there are significant differences in the coding among the different coders, we would like to understand the consequence of this. Since coder is the sole categorical variable in these models, it shifts the intercepts for the videos coded on PC 1 and PC 2, and the residuals provide proxy scores for the videos independent of coder. These can also be transformed back to the original coding scale for each variable by rotating them with the PCA variable loadings, adjusting for the mean and standard deviation of the original variables, and setting a threshold of 0.5 to clip the values to 0 and 1. This gives a predicted consensus coding for each video across coders which can be checked against the original codes for consistency. When this is done, we end up with the differences for each code in Table \ref{pairwise}. The column labels \textbf{f$\rightarrow$f}, \textbf{f$\rightarrow$t}, etc. indicate cases where the original coding of f corresponds to the predicted consensus coding of f or t, etc.

\begin{table}[h!]
  \begin{center}
    \label{tab:tablez1}
    \begin{tabular}{lrrrr} 
      \textbf{var} &\textbf{f$\rightarrow$f} & \textbf{f$\rightarrow$t} & \textbf{t$\rightarrow$f} & \textbf{t$\rightarrow$t} \\
      \hline
      v & 1255 & 21 & 33 & 3892 \\
      p & 3353 & 42 & 61 & 1745 \\
      n & 2937 & 30 & 19 & 2215 \\
      q & 3744 & 46 & 139 & 1242 \\
      cx & 2637 & 458 & 736 & 1378 \\
    \end{tabular}
  \end{center}
  \caption{\label{pairwise}Pairwise comparisons of corrected and original codes.}
\end{table}

The corrected \textbf{v}, \textbf{p}, and \textbf{n} are very close to their original values, \textbf{q} and \textbf{cx} less so. In Table \ref{percoder}, we summarize the corrections in terms of the individual coders, collapsing the values for \textbf{f$\rightarrow$f} and \textbf{t$\rightarrow$t} into a single row for each coder labeled \textbf{=}. As before, we find small numbers of corrections (predicted inter-coder differences) across variables \textbf{v}, \textbf{n} and \textbf{p}, and larger differences for \textbf{cx} and \textbf{q}. We conclude from this that we can safely use the original coding of \textbf{v}, \textbf{n} and \textbf{p} without alteration; in the subsequent diffusion analysis we do this by using PC2 without correction. Context panels (\textbf{cx}) require greater care in interpretation; coders coded different groups of dates, which could have different rates of deployment of the relevant context panels on YouTube. However, considering the greater interpretation required for \textbf{v}, \textbf{n} and \textbf{p} compared to the simple identification of a box existing in \textbf{cx}, this suggests that it is plausible that the differences in coding \textbf{cx} stem from the dataset rather than coding errors.

\begin{table}[h!]
  \begin{center}
    \label{tab:tablez2try}
    \begin{tabular}{ccrrrrr} 
    & $\Delta$ & \textbf{v} & \textbf{p} & \textbf{n} & \textbf{q} & \textbf{cx}\\
      \hline
1& f$\rightarrow$ t & 14 & 36 & 27 & 15 & 231 \\
& t$\rightarrow$ f & 3 & 37 & 6 & 47 & 404 \\
& = & 2716 & 2660 & 2700 & 2671 & 2098 \\
      \hline
2& f$\rightarrow$ t & 0 & 0 & 0 & 23 & 96 \\
& t$\rightarrow$ f & 4 & 6 & 2 & 18 & 139 \\
& = & 1311 & 1309 & 1313 & 1274 & 1080 \\
      \hline
3& f$\rightarrow$ t & 7 & 6 & 3 & 8 & 131 \\
& t$\rightarrow$ f & 26 & 18 & 11 & 74 & 193 \\
& = & 1120 & 1129 & 1139 & 1071 & 829 \\
    \end{tabular}
  \end{center}
  \caption{\label{percoder}Per-coder change comparisons ($\Delta$) between original and corrected variables.}
\end{table}

\section{Linking Clusters}
\label{sec:linking}
\begin{figure}
    \centering
    \includegraphics[width=1.0\linewidth]{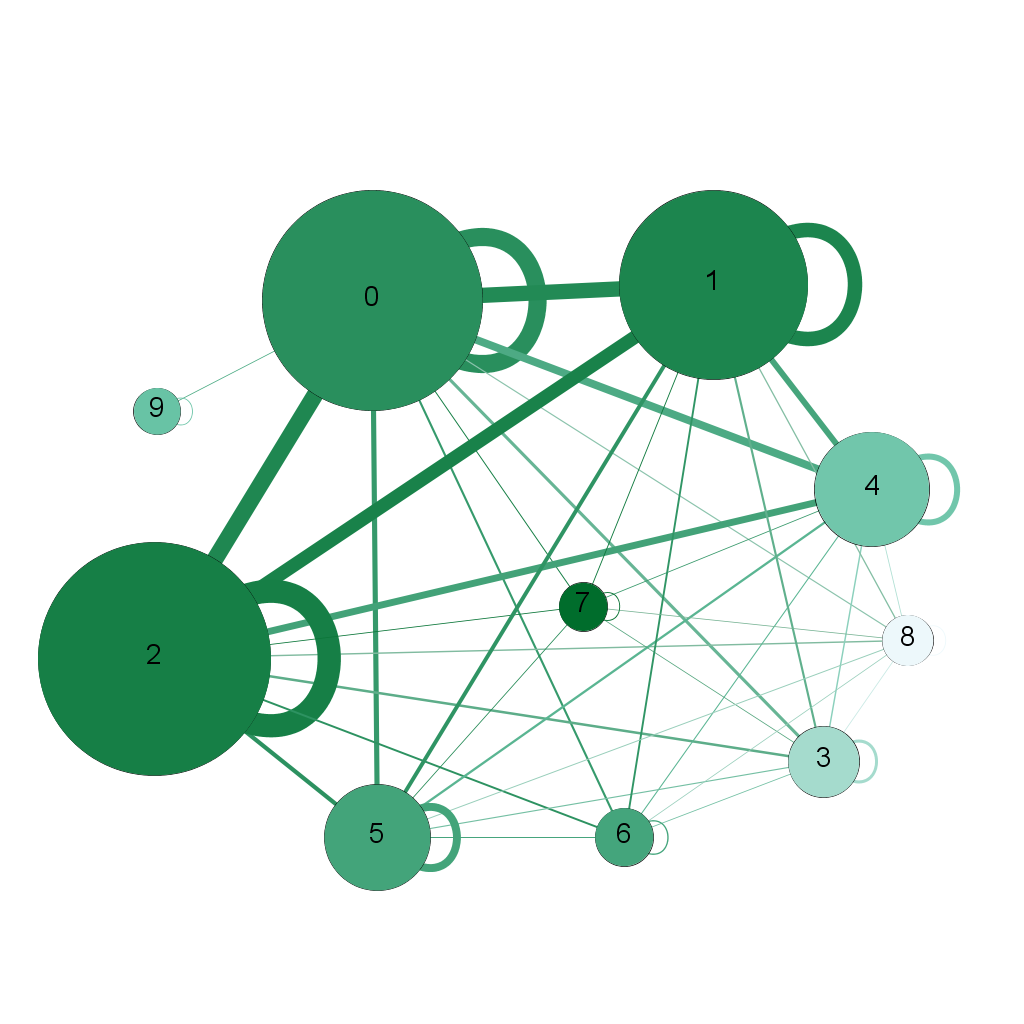}
    \caption{10 clusters represented by meta-nodes with edges and self-loops representing links between clusters and internal linking, respectively. Clusters were determined by applying the Louvain community detection algorithm on the links-links projection of the links-video network. Node color brightness encodes average loading of constituent nodes onto PC2.}
    \label{fig:10_cluster_net_pc2}
\end{figure}

Finally, we constructed a cluster analysis according to content (video) diffusion paths. To start we construct a user-link bipartite network where links are represented by the original shortened URLs. Shortened URLs can be generated by the system many times per video and so index a specific sharing path for a video. We apply the Louvain community detection algorithm \cite{python_louvain} to the link-link projection of 88,958 nodes and 9,043,668 edges, arriving at 2,305 community solution. These communities are then treated as meta-nodes and used these nodes to construct a community-video bipartite network, on which we again apply the Louvain algorithm, arriving at 10 clusters of diffusion communities that share videos in common between them. Figure \ref{fig:10_cluster_net_pc2} shows these 10 clusters as meta-nodes with edges between cluster and self loops weighted according to the number of links between and within a cluster, respectively. The average loading of the cluster's constituent nodes onto PC2 from the content analysis is mapped as the gradient color, with darker shades being more toward the loading of \textbf{n} and lighter being more toward the loading of \textbf{p}. Table \ref{tab:clustitles} lists the top three most-shared videos among each of these clusters.

Cluster 0, 1, 2, 4, and 5 dominate interaction within this model. Cluster 0  is characterized by mixed anti-vaccine content from multiple sources that otherwise overlaps with that of other clusters. This cluster represents the sharing behavior of twitter users who are not especially tied to particular domains within vaccine discourse beyond being broadly anti-vaccine. Cluster 1 contains YouTube videos from content creators who try to present themselves as not anti-vaccine despite producing content that is regularly consumed by anti-vax audiences. \href{https://www.youtube.com/c/thejimmydoreshow}{The Jimmy Dore Show} and \href{https://www.youtube.com/c/Campbellteaching}{Dr. John Campbell} are representative of this cluster. Cluster 2 predominantly contains content related to \href{https://www.youtube.com/c/veritasvisuals}{Project Veritas}, a channel premised on exposing corporate or left-wing "conspiracies." This content is very clearly anti-vaccine, but from conspiratorial perspectives. 

Cluster 3 contains largely the channel \href{https://www.youtube.com/c/WORKOUTSOLUTIONS/videos}{Workout Solutions Health Fitness}, which produces generally benign content on YouTube but employs a novel Twitter strategy of frequent self-promotion with high anti-vaccine messages. Cluster 4 contains more politically-minded anti-vaccine figures, like \href{https://www.youtube.com/c/RussellBrand}{Russel Brand} or \href{https://www.youtube.com/user/PrisonPlanetLive}{Paul Joseph Watson}. Cluster 5 includes doctors or other experts arguing against some part of pandemic public health guidelines. These doctors, such as Dr. Geert Vanden Bossche or Dr. Peter McCullough, are interviewed on many different YouTube channels. Cluster 6 contains mostly news reports that are often pro-vaccination or neutral on their own, but the tweets linking to them are at times arguing against these videos. Cluster 7 contains videos discussing vaccine or virus-related topics in technical detail that are not consistently pro or anti vaccine. Cluster 8 contains videos from \href{https://www.youtube.com/c/7news}{7NewsAustralia}, an Australian news broadcaster who consistently promotes vaccination. Cluster 9 consists of Christian fundamentalist antivax narratives, largely from the channel \href{https://www.youtube.com/c/AVoiceInTheDesert}{A Voice in the Desert}.

Cluster 1, 2, 4, and 5 can be seen as different archetypes of popular anti-vaccine discourse in our dataset, with 1 representing a more mainstream hesitancy, 2 representing conspiratorial discourse, 4 representing political discourse intersecting with anti-vaccine messages, and 5 representing the fusion of apparent expertise with anti-vaccine messages. These are held together by Cluster 0, representing the Twitter sharing of videos by people who sharing broadly anti-vaccination messages. 

\begin{table*}[!t]
  \begin{center}
    \begin{tabular}{rrl} % {p{0.5cm} p{1cm} p{14cm}}
    \\
Cl\# & Links & Video Title (truncated) \\
\hline
0 & 580 & Mass Vaccination in a Pandemic - Benefits versus Risks: Interview with Geert... \\
 & 252 & Vaccines and Related Biological Products Advisory Committee – 9/17/2021 \\
 & 239 & Spike protein inside nucleus enhancing DNA damage? - COVID-19 mRNA... \\
\hline
1 & 78 & EXPLOSIVE Truth About Vaccines \& COVID w/Inventor Of mRNA Vaccine... \\
 & 50 & Spike protein inside nucleus enhancing DNA damage? - COVID-19 mRNA vaccines... \\
 & 50 & Kyle's vaccine complication \\
\hline
2 & 157 & Carnicom Institute Disclosure Project - Overview with Clifford Carnicom \\
 & 104 & Pfizer Scientists: ‘Your [COVID] Antibodies Are Better Than The [Pfizer]... \\
 & 100 & Johnson \& Johnson: 'Kids Shouldn’t Get A F*cking [COVID] Vaccine;'... \\
\hline
3 & 67 & Vaccines and Related Biological Products Advisory Committee – 9/17/2021 \\
 & 55 & Back to School Morning Outdoor Weighted Cardio Rogue Fitness Fat Boy Sled \\
 & 53 & Bring Back DDT \& Dr Conover's Antibiotics For Wolbachia Co-Infections \\
\hline
4 & 52 & Vaccine Passports: THIS Is Where It Leads \\
 & 51 & Million March for Freedom Rally - London \\
 & 42 & Krystal Ball: Bill Gates Is LYING TO YOU On Vaccine Patent Protection \\
\hline
5 & 122 & Mass Vaccination in a Pandemic - Benefits versus Risks: Interview with Geert... \\
 & 29 & \#ScreenB4Vaccine: An Interview between Hooman Noorchashm MD, PhD and... \\
 & 25 & Peter McCullough, MD testifies to Texas Senate HHS Committee \\
\hline
6 & 34 & Vaccines and Related Biological Products Advisory Committee - 10/22/2020 \\
 & 22 & Eric Clapton: CANCELLED for exposing COVID-19 Vaccine \\
 & 10 &'Natural Immunity' Lawsuit Over COVID-19 Vaccine Mandate Ends in Surprising \\
\hline
7 & 18 & The Inventor of mRNA Vaccine Technology: Dr Robert Malone \\
 & 5 & Elon Musk on mRNA "You could turn someone into a freaking butterfly with the... \\
 & 2 & Italy Lawmaker Cunial Demands Arrest of "Vaccine Criminal" Bill Gates \\
\hline
8 & 20 & Rare inside tour of German lab creating mass Pfizer COVID-19 vaccines from... \\
 & 16 & Prime Minister Scott Morrison among the first Australians to receive COVID-19... \\
 & 14 & ATAGI recommends Pfizer vaccine be offered to Australian children as young as 12... \\
\hline
9 & 7 & Jesus and the Mark of the Beast \\
 & 4 & The Scientific Method And Jesus \\
 & 4 & The Truth About Christianity That Nobody Tells You \\
    \end{tabular}
  \end{center}
  \caption{Cluster number (Cl\#) with the top 3 linked videos (listed by their titles).}
  \label{tab:clustitles}
\end{table*}

\section{Discussion and Conclusion}
\label{sec:conc}
We began our investigation by sampling tweets at the Twitter-YouTube interface among users affected by vaccine-related moderation, among whom there is a propensity of anti-vaccine and anti-authority messages. Some use of pro-vaccine videos is made, but typically in refutation; other pro-vaccine videos are pushed to the edges of the discourse. Hence, we observe that the discursive impetus which led to moderating anti-vaccine information in the first place continues in spite of those efforts. Sharing specific anti-vaccination information may become harder under moderation, but the same users will often still share closely-related content.

Anti-vaccine discourse appears to persist through a predominantly political framing. Well-known figures taking anti-authoritarian stances on public health measures provide a framework into which dissident researchers, doctors, nurses, etc. can slot topically-relevant messages. These messages sow further doubt, and thereby these actors potentially rise to a new kind of prominence. John Campbell, Peter McCullough and Geert Vanden Bosche are three examples of such professionals whose public communication profiles have been raised as vaccination skeptics. Anti-vaccine content further differentiates along a political-medical axis among the users sharing it. 

At times, the connection of video content to vaccination is more tangential, e.g., in the case of \href{https://www.youtube.com/c/WORKOUTSOLUTIONS/videos}{Workout Solutions} (cluster 3), whose Twitter account contributed an astounding 30,417 tweets to our dataset, only 6,920 of which were retweets; the next nearest user had only 2,916 tweets. These tweets were highly formulaic, consisting of recombined phrases along with numerous hashtags and mentions for Canadian political figures (the account owner and related YouTube channel are in Ontario). Here, it appears that someone attempted to draw attention to his YouTube channel, possibly unsuccessfully,  by linking to ongoing anti-lockdown/anti-vaccination political discourses using some kind of third-party Twitter app. 

It also appears that moderation at the interface between YouTube and Twitter is not entirely successful. In part, this can be attributed to failures to fully implement the moderation stances adopted by the individual platforms. One such failure is the tendency for pro-vaccine messages to be flagged by context bars on YouTube while misinforming content is left un-flagged. It is possible that the means by which the context bars are applied is too simplistic, e.g. keyword searches in titles, descriptions and transcripts that are readily evaded by mutating the keywords. Alternative means, e.g. tracking known misinforming personalities, would potentially be much more effective, and more closely resembles how editorial discretion is exercised in print and broadcast media, but may aggravate such figures' dissident political stances. 

Regardless, the policy differential between Twitter and YouTube \emph{vis a vis} COVID-19 vaccination causes leakage. Other strategies at the disposal of a platform such as YouTube, such as search down-ranking of videos, are entirely side-stepped by cross-platform linking, where the original platform no longer has control over the spread of the content. Consequently, there appears to be a need for closer inter-platform cooperation in establishing and implementing moderation policies, although this too is likely to encounter political resistance from people who believe that either platform might unfairly moderate their messages.  

Our cross-platform approach has a number of limitations. Identifying topics through term clusters highlighted spam strategies used on Twitter. With respect to YouTube, video deletion for reasons other than terms of service violations raised further issues. For example, pro-vaccine videos from Indian state media entered into our sample due to the channel's decision to delete certain livestreams, potentially raising the number of pro-vaccine videos in our dataset. Similarly, videos hosted on YouTube but marked as "unlisted" may be found through Twitter; the reasons that content creators delist their videos are not necessarily consistent or clear. More generally, Twitter and YouTube users can always use the platform in unusual ways, and more careful culling from broad queries like that of CoVaxxy might be needed. Future work could address some of these data collection concerns, especially if differentiating the reason for the deletion of the original YouTube videos could be ascertained. This is not yet provided for by the YouTube API. Alternatively, future work could investigate this and similar datasets through more detailed content analysis approaches. 

\section*{Acknowledgments}

This work is supported in part by the Knight Foundation and the National Science Foundation (grant ACI-1548562). We also want to thank John Bryden and the rest of the CoVaxxy team for facilitating access to CoVaxxy data and Kai-Cheng Yang for helpful feedback.

\bibliography{acl2022}
\bibliographystyle{acl_natbib}

\end{document}